\documentclass[11pt,ams]{article}%
\usepackage{geometry}
\usepackage{dsfont}
\usepackage{amsmath}
\usepackage{amsfonts}
\usepackage{amssymb}
\usepackage{graphicx}%
\usepackage{subfig}
\usepackage{float}

\begin{document}

\date{}
%\title{\textbf{Gribov propagator and $i$-particles at work: constructing local composite operators for  glueball states }}
\title{\textbf{Constructing local composite operators for  glueball states from a confining Gribov propagator}}
\author{\textbf{M.~A.~L.~Capri}$^{a}$\thanks{capri@ufrrj.br}\,\,,
\textbf{A.~J.~G\'{o}mez}$^{b}$\thanks{ajgomez@uerj.br}\,\,,
\textbf{M.~S.~Guimaraes}$^{b}$\thanks{msguimaraes@uerj.br}\,\,,
\textbf{V.~E.~R.~Lemes}$^{b}$\thanks{vitor@dft.if.uerj.br}\,\,,\\
\textbf{S.~P.~Sorella}$^{b}$\thanks{sorella@uerj.br}\ \thanks{Work supported by
FAPERJ, Funda{\c{c}}{\~{a}}o de Amparo {\`{a}} Pesquisa do Estado do Rio de
Janeiro, under the program \textit{Cientista do Nosso Estado}, E-26/101.578/2010.}\,\,,\,\,\textbf{D.~G.~Tedesco}$^{b}$\thanks{dgtedesco@uerj.br}\,\,\\[2mm]
\textit{{\small $^{a}$ UFRRJ $-$ Universidade Federal Rural do Rio de Janeiro}}\\
\textit{{\small Departamento de F\'{\i}sica $-$ Grupo de F\'{\i}sica Te\'{o}rica e Matem\'{a}tica F\'{\i}sica}}\\
\textit{{\small BR 465-07, 23890-971, Serop\'edica, RJ, Brasil.}}\\
\textit{{\small {$^{b}$ UERJ $-$ Universidade do Estado do Rio de
Janeiro}}}\\\textit{{\small {Instituto de F\'{\i}sica $-$
Departamento de F\'{\i}sica Te\'{o}rica}}}\\\textit{{\small {Rua
S{\~a}o Francisco Xavier 524, 20550-013 Maracan{\~a}, Rio de
Janeiro, RJ, Brasil.}}}$$}
\maketitle
\begin{abstract}
\noindent The construction of BRST invariant local operators with the quantum numbers of the lightest glueball states, $J^{PC}= 0^{++}, 2^{++}, 0^{-+}$, is worked out by making use of an Euclidean  confining renormalizable gauge theory. The correlation functions of these operators are evaluated by employing a confining gluon propagator of the Gribov type and shown to display a spectral representation with positive spectral densities. An attempt to provide a first qualitative analysis of the ratios of the masses of the lightest glueballs is also discussed
\end{abstract}

\baselineskip=13pt

\newpage

%%%%%%%%%%%%%%%%%%%%%%
\section{Introduction}
%%%%%%%%%%%%%%%%%%%%%%
In a previous work \cite{Sorella:2010it},  we have outlined the construction of an Euclidean nonabelian gauge model, called the replica model,  in order to study aspects of the gluon confinement. As the name let it understand, the model relies on the introduction of a replica of the usual Faddeev-Popov action in the Landau gauge, where the gauge fields are coupled through a soft term, {\it i.e.} through a term which is quadratic in the fields.  The model  turns out to have many aspects in common with the Gribov-Zwanziger theory \cite{Gribov:1977wm,Zwanziger:1988jt,Zwanziger:1989mf}, a feature which enables us to employ it  in order to improve our current understanding of the role played by the Gribov horizon as well as of its consequences on the analytic properties of the correlation functions of gauge invariant local composite operators. \\\\As reported in \cite{Sorella:2010it}, the replica model enjoys the following properties
\begin{itemize}
\item the gluon propagator turns out to be a confining propagator of the Gribov type, namely
\begin{equation}
\langle A^a_\mu(k) A^b_\nu(-k) \rangle = \delta^{ab} \frac{k^2}{k^4+2\vartheta^4} \left( \delta_{\mu\nu} -\frac{k_\mu k_\nu}{k^2} \right) \;, \label{paai}
\end{equation}
where the mass parameter $\vartheta^2$ plays a role akin to that  of the Gribov mass parameter $\gamma^2$ of the Gribov-Zwanziger theory\footnote{See ref.\cite{Dudal:2009bf} for a general introduction to the  Gribov-Zwanziger theory and its  refined version.}
\item as the Gribov-Zwanziger theory, the replica model also displays a soft breaking of the BRST symmetry
\item it exhibits a nice interpretation in terms of $i$-particles \cite{Baulieu:2009ha}, {\it i.e.} in terms of the two unphysical modes with  complex conjugate imaginary masses $\pm i{\sqrt{2} \vartheta}^2$ corresponding to a confining Gribov propagator
\begin{equation}
\frac{k^2}{k^4+2\vartheta^4} = \frac{1}{2} \left( \frac{1}{k^2-i\sqrt{2}\vartheta^2} + \frac{1}{k^2+i\sqrt{2}\vartheta^2} \right) \;. \label{ip}
\end{equation}
As shown in \cite{Baulieu:2009ha}, the introduction of the $i$-particles provides a powerful tool in order to construct examples of local composite operators whose correlation functions do have good analyticity properties, as expressed by the K\"all\'{e}n-Lehmann spectral representation, and this in spite of the fact that they are evaluated by using a Gribov type propagator, eq.\eqref{paai},
\item analogously to the case of the Gribov-Zwanziger theory   \cite{Dudal:2007cw,Dudal:2008sp}, the replica model admits a refined version \cite{Sorella:2010it}, giving rise to a decoupling type gluon propagator which does not vanish at the origin in momentum space, $k^2=0$, {\it i.e.}
\begin{equation}
\langle A^{a}_{\mu}(k)A^{b}_{\nu}(-k) \rangle = \frac{k^{2}+m^{2}}{(k^{2}+m^{2})^{2}+2\vartheta^{4}}\delta^{ab}
\Bigl(\delta_{\mu\nu}-\frac{k_{\mu}k_{\nu}}{k^{2}}\Bigr)\,. \label{dec-prop}
\end{equation}
\end{itemize}
In the present paper we pursue the analysis of the replica model. We prove that the model is renormalizable to all orders and we discuss the construction of a set of local BRST invariant composite operators whose correlation functions exhibit a spectral representation.  We focus, in particular, on the BRST invariant operators having the quantum numbers of the three lightest glueball states\footnote{See ref.\cite{Mathieu:2008me} for a recent review  on glueballs.}, namely, $J^{PC}= 0^{++}, 2^{++}, 0^{-+}$. The corresponding spectral densities are worked out at one-loop order and shown to be positive.  A first qualitative analysis of the ratios of the masses  $m^2_{0^{++}}$, $m^2_{2^{++}}$, $m^2_{0^{-+}}$ is also presented by relying on  a phenomenological  SVZ-type sum rules\footnote{See ref.\cite{Colangelo:2000dp} for a pedagogical introduction to the SVZ sum rules and their applications to QCD.}  \cite{Shifman:1978bx,Shifman:1978by,Shifman:1979if,Novikov:1981xi} from which an expression for the glueball masses is established through the use of the Borel transformation. \\\\The work is organized as follows. In Sect.2 the main features of the replica model are briefly reviewed. In Sect.3 we derive the set of Ward identities fulfilled by the model and we establish its renormalizability to all orders by using the algebraic renormalization \cite{Piguet:1995er}. These results are proven to extend to the refined version of the model. Also, a simple argument accounting for the non-renormalization of the mass parameter $\vartheta^2$ appearing in the Gribov propagator, eq.\eqref{paai}, is provided. Sect.4. deals with the construction of local BRST invariant operators with the quantum numbers of the lightest glueball states $0^{++}, 2^{++}, 0^{-+}$. The relationship of these operators with the $i$-particles is discussed and the K\"all\'{e}n-Lehmann spectral representations for the corresponding correlation functions is established at one-loop order.  A qualitative analysis of the ratios of the masses  $m^2_{0^{++}}$, $m^2_{2^{++}}$, $m^2_{0^{-+}}$  is reported in Sect.5. Finally, Sect.6. gathers our conclusion.

%%%%%%%%%%%%%%%%%%%%%%%%%%%%%%%%%%%%%%%%%%%%%%%%%%%%%%%%%
\section{A short survey on the replica model}  \label{gm}
%%%%%%%%%%%%%%%%%%%%%%%%%%%%%%%%%%%%%%%%%%%%%%%%%%%%%%%%%
One starts by considering the Faddeev-Popov action in the Landau gauge
\begin{equation}
S_{{FP}} = \int d^4x\, \left(  \frac{1}{4} F^a_{\mu\nu}(A) F^a_{\mu\nu}(A) +i b^a\, \partial_\mu A^a_\mu +
{\bar c}^a \partial_{\mu} D^{ab}_\mu(A) c^b \right)  \;, \label{fpact}
\end{equation}
where $b^a$ is the Lagrange multiplier enforcing the Landau gauge condition, $\partial_{\mu} A^a_{\mu}=0$, $({\bar c}^a, c^a)$ stand for the Faddeev-Popov ghosts, $F^a_{\mu\nu}(A)$ is the field strength
\begin{equation}
F^a_{\mu\nu}(A) = \partial_{\mu} A^a_{\nu} - \partial_{\nu} A^a_{\mu} + g f^{abc} A^b_{\mu} A^c_{\nu} \;, \label{fst}
\end{equation}
and $D^{ab}_\mu(A) $ denotes the covariant derivative
\begin{equation}
D^{ab}_\mu(A) c^b = \partial_\mu c^a - g f^{abc} A^c_\mu c^b \;. \label{cda}
\end{equation}
The action of the replica  model is obtained as follows:
\begin{itemize}
\item one first considers a replica of the Faddeev-Popov action, eq.\eqref{fpact}, by introducing a set of mirror fields $(U^a_\mu, {\bar b}^a, {\bar \omega}^a, \omega^a)$ as well as a mirror Faddeev-Popov action:
\begin{equation}
S_{{MFP}} = \int d^4x\, \left(  \frac{1}{4} U^a_{\mu\nu}(U) U^a_{\mu\nu}(U) + i{\bar b}^a\, \partial_\mu U^a_\mu +
{\bar \omega}^a \partial_{\mu} D^{ab}_\mu(U) \omega^b \right)  \;, \label{mfpact}
\end{equation}
where $U^a_{\mu\nu}(U)$ is the field strength corresponding to the field $U^a_\mu$
\begin{equation}
U^a_{\mu\nu}(U) = \partial_{\mu} U^a_{\nu} - \partial_{\nu} U^a_{\mu} + g f^{abc} U^b_{\mu} U^c_{\nu} \;, \label{mfst}
\end{equation}
while ${\bar b}^a$ and $({\bar \omega}^a, \omega^a)$ stand for the mirror Lagrange multiplier and mirror Faddeev-Popov ghosts, respectively, and
\begin{equation}
D^{ab}_\mu(U) \omega^b = \partial_\mu \omega^a - g f^{abc} U^c_\mu \omega^b \;. \label{mcda}
\end{equation}
\item the gauge field $A^a_\mu$ is softly coupled to its mirror field $U^a_\mu$ through the following mixed term
\begin{equation}
S_{\vartheta} = i \sqrt{2} \vartheta^2 \int d^4x\; A^a_\mu U^a_\mu \;,  \label{mass}
\end{equation}
where, as already mentioned,  $\vartheta^2$ is a mass parameter which plays a role analogous to that  of the Gribov parameter $\gamma^2$ of the Gribov-Zwanziger action \cite{Gribov:1977wm,Zwanziger:1988jt,Zwanziger:1989mf}.
\end{itemize}
The replica model is thus specified by the following action
\begin{eqnarray}
S_{replica} & = & S_{{FP}} + S_{{MFP}} + S_{\vartheta}  \nonumber \\
    & = &  \int d^4x\, \biggl(  \frac{1}{4} F^a_{\mu\nu}(A) F^a_{\mu\nu}(A) + \frac{1}{4} U^a_{\mu\nu}(U) U^a_{\mu\nu}(U)
     + i \sqrt{2} \vartheta^2 A^a_\mu U^a_\mu \nonumber \\
    &&  + ib^a\, \partial_\mu A^a_\mu + {\bar c}^a \partial_{\mu} D^{ab}_\mu(A) c^b
    + i {\bar b}^a\, \partial_\mu U^a_\mu + {\bar \omega}^a \partial_{\mu} D^{ab}_\mu(U) \omega^b \biggr)  \;. \label{model}
\end{eqnarray}
Let us have a look at the propagators of the fields $(A^a_{\mu}, U^a_\mu)$, {\it i.e.}
\begin{equation}
\langle A^a_\mu(k) A^b_\nu(-k) \rangle = \delta^{ab} \frac{k^2}{k^4+2\vartheta^4} \left( \delta_{\mu\nu} -\frac{k_\mu k_\nu}{k^2} \right) \;, \label{paa}
\end{equation}
\begin{equation}
\langle U^a_\mu(k) U^b_\nu(-k) \rangle = \delta^{ab} \frac{k^2}{k^4+2\vartheta^4} \left( \delta_{\mu\nu} -\frac{k_\mu k_\nu}{k^2} \right) \;, \label{puu}
\end{equation}
\begin{equation}
\langle A^a_\mu(k) U^b_\nu(-k) \rangle = \delta^{ab} \frac{-i\sqrt{2}\vartheta^2}{k^4+2\vartheta^4} \left( \delta_{\mu\nu} -\frac{k_\mu k_\nu}{k^2} \right) \;. \label{pau}
\end{equation}
As one sees from expressions \eqref{paa}, \eqref{puu}, \eqref{pau}, all propagators are of the confining Gribov type. As such, they correspond to unphysical excitations. Said otherwise, we cannot attach a particle interpretation to the propagators of the elementary fields $(A^a_{\mu}, U^a_\mu)$. \\\\Another feature displayed by the model is that it contains a unique coupling constant $g$. Both fields $A^a_{\mu}$ and $U^a_\mu$ interact  with the same coupling. We remark that the feature of having a unique coupling constant is protected by a powerful discrete mirror symmetry \cite{Sorella:2010it}. It turns out in fact that the action $S_{replica}$, eq.\eqref{model}, is left invariant by the following discrete transformations
\begin{eqnarray}
A^a_{\mu}  \rightarrow  U^a_{\mu} \;, \nonumber \\
U^a_{\mu}  \rightarrow A^a_{\mu} \;, \nonumber \\
b^a  \rightarrow {\bar b}^a \;, \nonumber \\
{\bar b}^a  \rightarrow { b}^a \;, \nonumber \\
c^a  \rightarrow {\omega}^a \;, \nonumber \\
\omega^a  \rightarrow {c}^a \;, \nonumber \\
{\bar c}^a  \rightarrow {\bar \omega}^a \;, \nonumber \\
{\bar \omega}^a  \rightarrow {\bar c }^a \;. \label{mirror}
\end{eqnarray}
The mirror symmetry \eqref{mirror} means that the fields $(A^a_\mu, {b}^a, {\bar c}^a, c^a)$ can be replaced by  $(U^a_\mu, {\bar b}^a, {\bar \omega}^a, \omega^a)$, and vice-versa. As in the case of the Gribov-Zwanziger theory  \cite{Dudal:2009bf}, the action \eqref{model} displays a softly broken BRST invariance. It turns out in fact that the nilpotent BRST transformations
\begin{eqnarray}
s A^a_\mu & = & - D^{ab}_\mu(A) c^b \;, \nonumber \\
s U^a_\mu & = & - D^{ab}_\mu(U) \omega^b \;, \nonumber \\
s c^a & = & \frac{g}{2} f^{abc} c^b c^c \;, \nonumber \\
s \omega^a & = & \frac{g}{2} f^{abc} \omega^b \omega^c \;, \nonumber \\
s {\bar c}^a & = & ib^a  \;, \nonumber \\
s b^a & = & 0 \;, \nonumber \\
s {\bar \omega}^a & = & i{\bar b}^a  \;, \nonumber \\
s {\bar b}^a & = & 0 \;, \label{brst}
\end{eqnarray}
leave the action $S_{replica}$ invariant up to soft terms proportional to the parameter $\vartheta^2$, {\it i.e.}
\begin{equation}
s S_{replica}  = \vartheta^2 \Delta_{break} \;, \label{break}
\end{equation}
where $\Delta_{break}$ is given by
\begin{equation}
\Delta_{break} = -i \sqrt{2} \int d^4x \; \left( U^a_\mu D^{ab}_\mu(A)c^b + A^a_\mu D^{ab}_\mu(U)\omega^b \right) \;. \label{bd}
\end{equation}
Being of dimension two in the fields, $\Delta_{break}$ is a soft breaking. Finally, as discussed in \cite{Sorella:2010it}, the replica model reduces to the Faddeev-Popov action when  $\vartheta^2=0$.

\subsection{Relationship with the $i$-particles}
In view of the construction of  the BRST invariant local composite operators with the quantum numbers  $0^{++}, 2^{++}, 0^{-+}$, it is worth to spend a few words on the relationship between the replica model and the $i$-particles \cite{Baulieu:2009ha}, {\it i.e.} the pair of unphysical modes with complex masses $\pm i \sqrt{2} \vartheta^2$ associated to a confining Gribov type propagator.  To this end, let us consider the quadratic part of the action $S_{replica}$ containing the two gauge fields $(A^a_\mu, U^a_\mu)$, namely
\begin{equation}
S_{quad} = \int d^4x\; \left( \frac{1}{2} A^a_\mu (-\partial^2) A^a_\mu + \frac{1}{2} U^a_\mu (-\partial^2) U^a_\mu +
i\sqrt{2} \vartheta^2 A^a_\mu U^a_\mu \right) \;, \label{qd}
\end{equation}
where we have already taken into account the Landau gauge conditions, $\partial_\mu A^a_\mu=0$ and $\partial_\mu U^a_\mu=0$. Expression \eqref{qd} can be cast in diagonal form by introducing the two field combinations
\begin{eqnarray}
\lambda^a_\mu & = & \frac{1}{\sqrt{2}} \left( A^a_\mu + U^a_\mu \right) \;,  \nonumber \\
\eta^a_\mu & = & \frac{1}{\sqrt{2}} \left( A^a_\mu - U^a_\mu \right) \;.  \label{ip}
\end{eqnarray}
Therefore
\begin{equation}
S_{quad} = \int d^4x\; \left( \frac{1}{2} \lambda^a_\mu (-\partial^2+i\sqrt{2}\vartheta^2) \lambda^a_\mu + \frac{1}{2} \eta^a_\mu (-\partial^2-i\sqrt{2}\vartheta^2) \eta^a_\mu   \right) \;, \label{qd}
\end{equation}
which describes in fact the propagation of two unphysical modes with complex masses $\pm i\sqrt{2}\vartheta^2$
\begin{eqnarray}
\langle \lambda^a_\mu(k) \lambda^b_\nu(-k) \rangle & = &  \frac{1}{2} \langle (A^a_\mu(k) + U^a_\mu(k) )
(A^b_\nu(-k) + U^b_\nu(-k)) \rangle =  \delta^{ab} \frac{1}{k^2+i\sqrt{2}\vartheta^2} \left( \delta_{\mu\nu} -\frac{k_\mu k_\nu}{k^2} \right) \;, \nonumber \\
\langle \eta^a_\mu(k) \eta^b_\nu(-k) \rangle & = &      \frac{1}{2} \langle (A^a_\mu(k) - U^a_\mu(k) )
(A^b_\nu(-k) - U^b_\nu(-k)) \rangle  =         \delta^{ab} \frac{1}{k^2-i\sqrt{2}\vartheta^2} \left( \delta_{\mu\nu} -\frac{k_\mu k_\nu}{k^2} \right) \;. \nonumber \\
\langle \lambda^a_\mu(k) \eta^b_\nu(-k) \rangle & = & \frac{1}{2} \langle (A^a_\mu(k) + U^a_\mu(k) )
(A^b_\nu(-k) - U^b_\nu(-k)) \rangle = 0 \;.
\label{ipprop}
\end{eqnarray}
These are precisely the $i$-particles corresponding to the Gribov propagators in eqs.\eqref{paa},  \eqref{puu}, \eqref{pau}. We see thus that the action \eqref{model} has a direct interpretation in terms of $i$-particles. As observed  in \cite{Baulieu:2009ha}, the advantage of introducing the fields $(\lambda^a_\mu, \eta^a_\mu)$ relies on the fact that they turn out to be  helpful in order to construct examples of local operators whose correlation functions exhibit the  K\"all\'{e}n-Lehmann spectral representation. This feature stems from the observation that the  momentum integrals corresponding to Feynman diagrams containing an equal number of  propagators of the $\lambda$-type and of the $\eta$-type can be cast in the form of a spectral representation, see  \cite{Baulieu:2009ha} for a detailed discussion. Let us consider in fact the one-loop integral
\begin{equation}
{\cal I}(k^2) = \int \frac{d^4p}{(2\pi)^4} \; \frac{1}{\left( (k-p)^2+i\sqrt{2}\vartheta^2 \right) \left(  p^2 -i \sqrt{2}\vartheta^2 \right)}  \;.  \label{bint}
\end{equation}
For the corresponding spectral representation one has  \cite{Baulieu:2009ha}
\begin{equation}
{\cal I}(k^2) - {\cal I}(0) = \int_{2\sqrt{2}\vartheta^2}^{\infty} d\tau \rho(\tau) \left( \frac{1}{\tau+k^2} -\frac{1}{\tau} \right)  \;, \label{iint}
\end{equation}
where the spectral density
\begin{equation}
\rho(\tau) = \frac{1}{16\pi^2} \frac{\sqrt{\tau^2-8\vartheta^4}}{\tau}  \;, \label{spectr}
\end{equation}
is positive in the range of integration\footnote{The subtraction of the factor ${\cal I}(0)$ in eq.\eqref{iint} is needed to account for the divergent character of expression \eqref{bint} in four dimensions.}. This relevant property enables us to construct local operators with good analyticity properties. As an example, we quote the operator
\begin{equation}
{\cal O}_{\lambda\eta} = \left( \partial_\mu \lambda^a_\nu - \partial_\nu \lambda^a_\mu \right) \left( \partial_\mu \eta^a_\nu - \partial_\nu \eta^a_\mu \right)  \;, \label{iop}
\end{equation}
also extensively investigated in \cite{Baulieu:2009ha}, where its two-point correlation function has been shown to be cast in the form of a spectral representation with positive spectral function\footnote{Also here, a suitable subtraction to get rid of ultraviolet divergences is needed, see \cite{Baulieu:2009ha} for details.}
\begin{eqnarray}
\langle {\cal O}_{\lambda\eta}(k) {\cal O}_{\lambda\eta}(-k) \rangle & = & \int_{2\sqrt{2}\vartheta^2}^{\infty} d\tau \frac{\rho(\tau)}{\tau+k^2}   \;, \nonumber \\
\rho(\tau) & =   & 12(N^2-1) \frac{ \sqrt{\tau^2-8\vartheta^4}\;(8\vartheta^4+\tau^2)}{32\pi^2\tau}
%\frac{4(N^2-1)}{\pi \sqrt{\tau^2-8\theta^4}}
 \;. \label{id}
\end{eqnarray}

%%%%%%%%%%%%%%%%%%%%%%%%%%%%%%%%%%%%%%%%%%%%%%%%%%%%%%%%%%%%%%
\section{Renormalizability of the replica model and non-renormalization properties of the mass parameter $\vartheta^2$}
%%%%%%%%%%%%%%%%%%%%%%%%%%%%%%%%%%%%%%%%%%%%%%%%%%%%%%%%%%%%%%
The existence of a soft breaking of the BRST symmetry does not prevent us to establish a set of Slavnov-Taylor identities which are suitable for an all-order algebraic analysis  of the renormalizability properties of the model. The usual way of proceeding\footnote{See, for example, the case of the Slavnov-Taylor identities derived in the Gribov-Zwanziger theory  \cite{Dudal:2009bf,Dudal:2008sp,Dudal:2010fq}.} is that of  introducing an extended action which incorporates all local composite operators entering the soft breaking, by coupling them to a suitable set of external sources. The original action is thus recovered when the  sources acquire a particular value, which we shall refer to as the physical value. The renormalizability of the extended action entails thus the renormalizability of the starting action $S_{replica}$, eq.\eqref{model}. \\\\The use of the so-called algebraic renormalization is particularly suited for the study of a field theory whose starting classical action is characterized by a set of symmetries, as in the case of the replica model considered here. For such theories the algebraic renormalization provides a purely algebraic set up in order to establish an all orders proof of the
renormalizability in a regularization independent way. Essentially, the algebraic renormalization combines in a powerful way the locality properties of the perturbative series, as expressed by the power-counting, with the set of Ward identities following from the symmetry content of the starting action. More precisely, from the power-counting,
one knows that the ultraviolet divergences which originate through radiative corrections are local polynomials in the fields and their derivatives whose dimensions are bounded by four. These local polynomials are thus completely characterized by the requirement that they have to be compatible with the set of Ward identities corresponding to the symmetries of the starting action. In much the same way, the power-counting ensures that  potential anomalies are also associated to local polynomials in the fields and their
derivatives which are constrained by the Wess-Zumino consistency conditions stemming from
the algebraic relations among the various functional operators associated to the Ward identities.
In summary, both invariant counterterms and potential anomalies can be fully characterized in
a purely algebraic way as the most general solution in the space of the local field polynomials
of the Ward identities which express the symmetry content of the model. This program goes
under the name of algebraic renormalization \cite{Piguet:1995er}.

%%%%%%%%%%%%%%%%%%%%%%%%%%%%%%%%%%%%%%%%%%%%%%%%
\subsection{Identifying the classical extended action}
%%%%%%%%%%%%%%%%%%%%%%%%%%%%%%%%%%%%%%%%%%%%%%%%
In order to write down the Slavnov-Taylor Ward identities,  we follow the general set up of the algebraic renormalization \cite{Piguet:1995er} and start by introducing a term depending on BRST invariant external sources coupled to the nonlinear BRST transformations of the fields appearing in eq.\eqref{brst}, namely
\begin{eqnarray}
S_{\mathrm{external}}&=&\int d^{4}x\,\Bigl[\Omega^{a}_{\mu}(sA^{a}_{\mu})+L^{a}(sc^{a})
+\bar{\Omega}^{a}_{\mu}(sU^{a}_{\mu})+\bar{L}^{a}(s\omega^{a})\Bigr]\nonumber\\
&=&\int d^{4}x\,\biggl(-\Omega^{a}_{\mu}\,D^{ab}_{\mu}(A)c^{b}+\frac{g}{2}f^{abc}L^{a}c^{b}c^{c}
-\bar{\Omega}^{a}_{\mu}\,D^{ab}_{\mu}(U)\omega^{b}+\frac{g}{2}f^{abc}\bar{L}^{a}\omega^{b}\omega^{c}\biggr)\,.
\end{eqnarray}
Notice that the mirror symmetry \eqref{mirror} can be immediately extended to the BRST sources by requiring that
\begin{eqnarray}
&\Omega^{a}_{\mu}\to\bar\Omega^{a}_{\mu}\,,\qquad\bar{\Omega}^{a}_{\mu}\to\Omega^{a}_{\mu}\,,&\nonumber\\
&L^{a}\to\bar{L}^{a}\,,\qquad\bar{L}^{a}\to L^{a}\,.&
\end{eqnarray}
We proceed now by introducing a BRST doublet \cite{Piguet:1995er} of external souces\footnote{As we shall see later,  the introduction of these sources with two color indices $(a,b)$ is useful in order to forbid the appearance of counterterms of the kind  $(A^{a}_{\mu}A^{a}_{\mu}+U^{a}_{\mu}U^{a}_{\mu})$.}
\begin{equation}
sK^{ab}=J^{ab}\,,\qquad sJ^{ab}=0\,,
\end{equation}
and obtain the following extended BRST invariant term
\begin{equation}
S_{J}=s\int d^{4}x\,K^{ab}A^{a}_{\mu}U^{b}_{\mu}=\int d^{4}x\,\Bigl[J^{ab}A^{a}_{\mu}U^{b}_{\mu}
+K^{ab}(D^{ac}_{\mu}(A)c^{c})U^{b}_{\mu}+K^{ab}A^{a}_{\mu}D^{bc}_{\mu}(U)\omega^{c}\Bigr]\,.
\end{equation}
As one can easily check, the soft  term $S_{\vartheta}$, eq.\eqref{mass}, can be recovered from  the invariant term $S_J$ when the external sources $J^{ab}$ and $K^{ab}$ attain the following physical values
\begin{equation}
J^{ab}\bigl|_{\mathrm{phys}}=i\sqrt{2}\vartheta^{2}\delta^{ab}\,,\qquad K^{ab}\bigl|_{\mathrm{phys}}=0\,,\qquad S_{J}\bigl|_{\mathrm{phys}}=S_{\vartheta}\,. \label{physv}
\end{equation}
Also, the  invariance under  the mirror symmetry  \eqref{mirror} is guaranteed by demanding that
\begin{equation}
J^{ab}\to J^{ba}\,,\qquad K^{ab}\to K^{ba}\,.
\end{equation}
Finally, for renormalization purposes,  an extra  BRST exact term has to be introduced
\begin{eqnarray}
S_{\mathrm{extra}}&=&\frac{\zeta}{2}s\int d^4x\,\Bigl(K^{ab}J^{ab}-gf^{abc}K^{ad}K^{bd}c^{c}-gf^{abc}K^{da}K^{db}\omega^{c}\Bigr)\nonumber\\
&=&\zeta\int d^{4}x\,\biggl(\frac{1}{2}\,J^{ab}J^{ab}-gf^{abc}J^{ad}K^{bd}c^{c}-gf^{abc}J^{da}K^{db}\omega^{c}
-\frac{g^{2}}{4}f^{abc}f^{cmn}K^{ad}K^{bd}c^{m}c^{n}\nonumber\\
&&-\frac{g^{2}}{4}f^{abc}f^{cmn}K^{da}K^{db}\omega^{m}\omega^{n}\biggr)\,.
\end{eqnarray}
Here, $\zeta$ is a constant parameter and, when the physical values of the sources \eqref{physv}  are taken, only the first term of the r.h.s. survives, giving rise to a constant vacuum term
\begin{equation}
S_{\mathrm{extra}}\bigl|_{\mathrm{phys}}=-\zeta(N^{2}-1)\gamma^{4}V\,,
\end{equation}
where $V$ is the 4-dimensional Euclidean volume.\\\\
Thus, the starting extended action we shall consider reads
\begin{equation}
\Sigma=S_{\mathrm{FP}}+S_{\mathrm{MFP}}+S_{\mathrm{external}}+S_{J}+S_{\mathrm{extra}}\label{Sigma}
\end{equation}

%%%%%%%%%%%%%%%%%%%%%%%%%%%%
\subsection{Ward identities}
%%%%%%%%%%%%%%%%%%%%%%%%%%%%
In oder to discuss  the renormalizability of the extended action $\Sigma$, eq.\eqref{Sigma}, let us give the full set of  Ward identities fulfilled by $\Sigma$. These are
\begin{itemize}
\item{the Landau gauge fixing equations \cite{Piguet:1995er}
\begin{equation}
\frac{\delta\Sigma}{\delta b^{a}}=i\partial_{\mu}A^{a}_{\mu}\,,\qquad
\frac{\delta\Sigma}{\delta \bar{b}^{a}}=i\partial_{\mu}U^{a}_{\mu}\,.  \label{gfix}
\end{equation}}
\item{the anti-ghosts Ward identities \cite{Piguet:1995er}
\begin{equation}
\frac{\delta\Sigma}{\delta\bar{c}^{a}}+\partial_{\mu}\frac{\delta\Sigma}{\delta{\Omega}^{a}_{\mu}}=0\,,\qquad
\frac{\delta\Sigma}{\delta\bar{\omega}^{a}}+\partial_{\mu}\frac{\delta\Sigma}{\delta{\bar\Omega}^{a}_{\mu}}=0\,.
\end{equation}}
\item{the Slavnov-Taylor identities \cite{Piguet:1995er}
\begin{equation}
{\mathcal{S}}(\Sigma)  = 0  \;,  \label{stw}
\end{equation}
with
\begin{eqnarray}
\mathcal{S}(\Sigma)=\int d^{4}x\,\biggl(
\frac{\delta\Sigma}{\delta{A}^{a}_{\mu}}\frac{\delta\Sigma}{\delta{\Omega}^{a}_{\mu}}
+\frac{\delta\Sigma}{\delta{U}^{a}_{\mu}}\frac{\delta\Sigma}{\delta{\bar\Omega}^{a}_{\mu}}
+\frac{\delta\Sigma}{\delta{c}^{a}}\frac{\delta\Sigma}{\delta{L}^{a}}
+\frac{\delta\Sigma}{\delta{\omega}^{a}}\frac{\delta\Sigma}{\delta{\bar{L}}^{a}}
+ib^{a}\frac{\delta\Sigma}{\delta{\bar{c}}^{a}}
+i\bar{b}^{a}\frac{\delta\Sigma}{\delta{\bar{\omega}}^{a}}
+J^{ab}\frac{\delta\Sigma}{\delta{K}^{ab}}\biggr)\,. \nonumber \\
\end{eqnarray}}
\item{the two rigid symmetries
\begin{eqnarray}
\mathcal{W}^{a}(\Sigma)&=&gf^{abc}\int d^{4}x\,\biggl(
A^{b}_{\mu}\frac{\delta\Sigma}{\delta{A}^{c}_{\mu}}
+b^{b}\frac{\delta\Sigma}{\delta{b}^{c}}
+\bar{c}^{b}\frac{\delta\Sigma}{\delta\bar{c}^{c}}
+c^{b}\frac{\delta\Sigma}{\delta{c}^{c}}
+\Omega^{b}_{\mu}\frac{\delta\Sigma}{\delta{\Omega}^{c}_{\mu}}\nonumber\\
&&+L^{b}\frac{\delta\Sigma}{\delta{L}^{c}}
+J^{bd}\frac{\delta\Sigma}{\delta{J}^{cd}}
+K^{bd}\frac{\delta\Sigma}{\delta{K}^{cd}}\biggr)=0\,,\label{rig}\\
\overline{\mathcal{W}}^{a}(\Sigma)&=&gf^{abc}\int d^{4}x\,\biggl(
U^{b}_{\mu}\frac{\delta\Sigma}{\delta{U}^{c}_{\mu}}
+\bar{b}^{b}\frac{\delta\Sigma}{\delta{\bar{b}}^{c}}
+\bar{\omega}^{b}\frac{\delta\Sigma}{\delta\bar{\omega}^{c}}
+\omega^{b}\frac{\delta\Sigma}{\delta{\omega}^{c}}
+\bar\Omega^{b}_{\mu}\frac{\delta\Sigma}{\delta{\bar\Omega}^{c}_{\mu}}\nonumber\\
&&+\bar{L}^{b}\frac{\delta\Sigma}{\delta{\bar{L}}^{c}}
+J^{db}\frac{\delta\Sigma}{\delta{J}^{dc}}
+K^{db}\frac{\delta\Sigma}{\delta{K}^{dc}}\biggr)=0\,.\label{rig2}
\end{eqnarray}
These symmetries imply that the first color index of the external sources $J^{ab}$ and $K^{ab}$ can be contracted only with the fields of the $A$-family, \emph{i.e.}
$(A^{a}_{\mu},b^{a},\bar{c}^{a},c^{a},\Omega^{a}_{\mu}, L^{a})$, while the second index can be contracted only with the fields of the $U$-family, $(U^{a}_{\mu},\bar{b}^{a},\bar{\omega}^{a},\omega^{a},\bar\Omega^{a}_{\mu}, \bar{L}^{a})$. They forbid, for example, the presence of counterterms  like $J^{ab}A^{a}_{\mu}A^{b}_{\mu}$ and
$J^{ab}U^{a}_{\mu}U^{b}_{\mu}$.}
\item{the two ghost Ward identities \cite{Blasi:1990xz,Piguet:1995er}
\begin{equation}
\mathcal{G}^{a}(\Sigma)=\Delta^{a}_{\mathrm{class}}\,,\qquad
\overline{\mathcal{G}}^{a}(\Sigma)=\bar\Delta^{a}_{\mathrm{class}}\,, \label{ghwid}
\end{equation}
where
\begin{eqnarray}
\mathcal{G}^{a}&=&\int d^{4}x\,\biggl(\frac{\delta}{\delta{c}^{a}}
-igf^{abc}\bar{c}^{b}\frac{\delta}{\delta{b}^{c}}
+gf^{abc}K^{bd}\frac{\delta}{\delta J^{cd}}\biggr)\,,\\
\Delta^{a}_{\mathrm{class}}&=&\int d^{4}x\,f^{abc}\biggl(\Omega^{b}_{\mu}A^{c}_{\mu}
-L^{b}c^{c}- g^2\zeta f^{amc}f^{dbq}K^{md}K^{cb}\omega^{q}\biggr)\,,
\end{eqnarray}
and
\begin{eqnarray}
\overline{\mathcal{G}}^{a}&=&\int d^{4}x\,\biggl(\frac{\delta}{\delta{\omega}^{a}}
-igf^{abc}\bar\omega^{b}\frac{\delta}{\delta{\bar{b}}^{c}}
+gf^{abc}K^{db}\frac{\delta}{\delta J^{dc}}\biggr)\,,\\
\bar\Delta^{a}_{\mathrm{class}}&=&\int d^{4}x\,f^{abc}\biggl(\bar\Omega^{b}_{\mu}U^{c}_{\mu}
-\bar{L}^{b}\omega^{c} - g^2\zeta f^{amc}f^{dbq}K^{dm}K^{bc}c^{q}\biggr)\,.
\end{eqnarray}}
Notice that the expressions  $\Delta^{a}_{\mathrm{class}}, \bar\Delta^{a}_{\mathrm{class}}$ in eq.\eqref{ghwid} are linear in the quantum fields. As such, they represent classical breakings, not affected by quantum corrections \cite{Piguet:1995er}.
\item{the $SL(2,\mathbb{R})$ symmetries
\begin{eqnarray}
\mathcal{D}(\Sigma)&=&\int d^{4}x\,\biggl(
c^{a}\frac{\delta\Sigma}{\delta\bar{c}^{a}}
-i\frac{\delta\Sigma}{\delta{b}^{a}}\frac{\delta\Sigma}{\delta{L}^{a}}\biggr)=0\,,\\
\overline{\mathcal{D}}(\Sigma)&=&\int d^{4}x\,\biggl(
\omega^{a}\frac{\delta\Sigma}{\delta\bar{\omega}^{a}}
-i\frac{\delta\Sigma}{\delta{\bar{b}}^{a}}\frac{\delta\Sigma}{\delta{\bar{L}}^{a}}\biggr)=0\,.  \label{sl2r}
\end{eqnarray}}
\end{itemize}

%%%%%%%%%%%%%%%%%%%%%%%%%%%%%%%%%%%%%%%%%
\subsection{Characterization of the most general invariant counterterm}
%%%%%%%%%%%%%%%%%%%%%%%%%%%%%%%%%%%%%%%%%
In order to characterize the most general invariant counterterm which can be
freely added to all orders in perturbation theory \cite{Piguet:1995er}, we
perturb the classical action $\Sigma$ by adding an integrated local
polynomial $\Sigma_c$ of dimension bounded by four, and with vanishing
ghost number. We demand thus that the perturbed
action, $(\Sigma+\eta\Sigma_c)$, where $\eta$ is an expansion
parameter, fulfills, to the first order in $\eta$, the same Ward
identities fulfilled by the classical action $\Sigma$. This requirement gives rise to the following
constraints for the counterterm $\Sigma_c$
\begin{eqnarray}
\frac{\delta}{\delta b^{a}}\Sigma_{c}&=&0\,,   \qquad  \frac{\delta}{\delta \bar{b}^{a}}\Sigma_{c}=0\,,\nonumber\\
\biggl(\frac{\delta}{\delta \bar{c}^{a}}
+\partial_{\mu}\frac{\delta}{\delta{\Omega}^{a}_{\mu}}\biggr)\Sigma_{c}&=&0\,,\qquad
\biggl(\frac{\delta}{\delta \bar{\omega}^{a}}
+\partial_{\mu}\frac{\delta}{\delta{\bar\Omega}^{a}_{\mu}}\biggr)\Sigma_{c}= 0\,,\nonumber\\
\mathcal{W}^{a}\Sigma_{c}&=&0\,,\qquad
\overline{\mathcal{W}}^{a}\Sigma_{c}=0\,,\nonumber\\
\mathcal{G}^{a}\Sigma_{c}&=&0\,,\qquad
\overline{\mathcal{G}}^{a}\Sigma_{c}=0\,,\nonumber\\
\mathcal{D}_{\Sigma}\Sigma_{c}&=&0\,,\qquad
\overline{\mathcal{D}}_{\Sigma}\Sigma_{c}=0\,, \nonumber\\
\mathcal{S}_{\Sigma}\Sigma_{c}&=&0\,,\nonumber\\
\label{constraits}
\end{eqnarray}
where $\mathcal{S}_{\Sigma}$, $\mathcal{D}_{\Sigma}$ and $\overline{\mathcal{D}}_{\Sigma}$ denote  the linearized operators
\begin{eqnarray}
\mathcal{S}_{\Sigma}&=&\int d^{4}x\,\biggl(
\frac{\delta\Sigma}{\delta{A}^{a}_{\mu}}\frac{\delta}{\delta{\Omega}^{a}_{\mu}}
+\frac{\delta\Sigma}{\delta{\Omega}^{a}_{\mu}}\frac{\delta}{\delta{A}^{a}_{\mu}}
+\frac{\delta\Sigma}{\delta{U}^{a}_{\mu}}\frac{\delta}{\delta{\bar\Omega}^{a}_{\mu}}
+\frac{\delta\Sigma}{\delta{\bar\Omega}^{a}_{\mu}}\frac{\delta}{\delta{U}^{a}_{\mu}}
+\frac{\delta\Sigma}{\delta{c}^{a}}\frac{\delta}{\delta{L}^{a}}
+\frac{\delta\Sigma}{\delta{L}^{a}}\frac{\delta}{\delta{c}^{a}}\nonumber\\
&&
+\frac{\delta\Sigma}{\delta{\omega}^{a}}\frac{\delta}{\delta{\bar{L}}^{a}}
+\frac{\delta\Sigma}{\delta\bar{L}^{a}}\frac{\delta}{\delta\omega^{a}}
+ib^{a}\frac{\delta}{\delta{\bar{c}}^{a}}
+i\bar{b}^{a}\frac{\delta}{\delta{\bar{\omega}}^{a}}
+J^{ab}\frac{\delta}{\delta{K}^{ab}}\biggr)\,,\\
\mathcal{D}_{\Sigma}&=&\int d^{4}x\,\biggl(
c^{a}\frac{\delta}{\delta\bar{c}^{a}}
-i\frac{\delta\Sigma}{\delta{b}^{a}}\frac{\delta}{\delta{L}^{a}}
-i\frac{\delta\Sigma}{\delta{L}^{a}}\frac{\delta}{\delta{b}^{a}}\biggr)\,,\\
\overline{\mathcal{D}}_\Sigma&=&\int d^{4}x\,\biggl(
\omega^{a}\frac{\delta\Sigma}{\delta\bar{\omega}^{a}}
-i\frac{\delta\Sigma}{\delta{\bar{b}}^{a}}\frac{\delta}{\delta{\bar{L}}^{a}}
-i\frac{\delta\Sigma}{\delta{\bar{L}}^{a}}\frac{\delta}{\delta{\bar{b}}^{a}}\biggr)\,.
\end{eqnarray}
Following the general set up of the algebraic renormalization \cite{Piguet:1995er}, it turns out that the most general invariant counterterm $\Sigma_{c}$ obeying the constraints \eqref{constraits} and the mirror symmetry is given by
\begin{eqnarray}
\Sigma_{c}&=&a_{0}\, \int d^4x \left( \frac{1}{4} F^a_{\mu\nu} F^a_{\mu\nu} + \frac{1}{4} U^a_{\mu\nu} U^a_{\mu\nu}\right) +\mathcal{S}_{\Sigma}\int d^{4}x\,\Bigl\{a_1\,\left((\Omega^{a}_{\mu}+\partial_{\mu}\bar{c}^{a})A^{a}_{\mu}
+(\bar\Omega^{a}_{\mu}+\partial_{\mu}\bar{\omega}^{a})U^{a}_{\mu}\right) \nonumber\\
&&+a_2\,K^{ab}A^{a}_{\mu}U^{b}_{\mu}
+\frac{a_3\zeta}{2}(K^{ab}J^{ab}
-gf^{abc}K^{ad}K^{bd}c^{c}-gf^{abc}K^{da}K^{db}\omega^{c})\Bigr\}\,.
\label{countert}
\end{eqnarray}
where $a_{0},a_1, a_2,a_3$ are free coefficients.
%%%%%%%%%%%%%%%%%%%%%%%%%%%%%%%%%%%%
\subsection{Renormalization factors}
%%%%%%%%%%%%%%%%%%%%%%%%%%%%%%%%%%%%
It remains now to show that the invariant counterterm \eqref{countert} can be reabsorbed through a redefinition of
the parameters, fields and sources of the classical starting action $\Sigma$, according to
\begin{equation}
\phi_0=Z^{1/2}_{\phi}\,\phi\,,\qquad\Phi_0=Z_{\Phi}\,\Phi\,,
\end{equation}
where
\begin{eqnarray}
\phi&\equiv&\{A,U,b, \bar{b}, c,\omega,\bar{c},\bar\omega\}\,,\nonumber\\
\Phi&\equiv&\{g,\zeta,\Omega,\bar\Omega,L,\bar{L},J,K\}\,,
\end{eqnarray}
so that
\begin{equation}
\Sigma(\phi_0,\Phi_0)=\Sigma(\phi,\Phi)+\eta\Sigma_{c}(\phi,\Phi)+O(\eta^2)\,.
\label{renorm}
\end{equation}
By direct inspection, the renormalization factors are found to be
\begin{eqnarray}
&Z_{U}^{1/2}=Z_{b}^{-1/2}=Z_{\bar{b}}^{-1/2}=Z^{1/2}_{A}\,,&\nonumber\\
&Z_{c}^{1/2}=Z_{\bar{c}}^{1/2}=Z_{\omega}^{1/2}=Z_{\bar{\omega}}^{1/2}=Z_{\Omega}=Z_{\bar\Omega}
=Z_{L}^{-1/2}=Z_{\bar{L}}^{-1/2}=Z_{g}^{-1/2}Z_{A}^{-1/4}\,,&\nonumber\\
&Z_{K}=Z_{J}Z_g^{-1/2}Z_{A}^{1/4}\,,&\label{Zs}
\end{eqnarray}
with
\begin{eqnarray}
Z_{A}^{1/2}&=&1+\epsilon\,\frac{a_0+2a_1}{2}\,,\nonumber\\
Z_{g}&=&1-\epsilon\,\frac{a_{0}}{2}\,,\nonumber\\
Z_{J}&=&1-\epsilon\,(a_0-a_2)\,,\nonumber\\
Z_{\zeta}&=&1+\epsilon\,(2a_0-2a_2+a_3)\,.\label{Zss}
\end{eqnarray}
Equations \eqref{renorm}, \eqref{Zs}, \eqref{Zss} show that the counterterm $\Sigma_{c}$ can be reabsorbed by means of a redefinition
of the fields, sources and parameters of the starting action $\Sigma$, establishing thus the renormalizability of
the replica model.

%%%%%%%%%%%%%%%%%%%%%%%%%%%%%%%%%%%%%%%%%%%
\subsection{Non-renormalization properties of the mass parameter $\vartheta^2$}
%%%%%%%%%%%%%%%%%%%%%%%%%%%%%%%%%%%%%%%%%%%
In this section we shall show, by graphical arguments, that the term
\begin{equation}
(J^{ab}A^{a}_{\mu}U^{a}_{\mu})\bigl|_{\mathrm{phys}}=i\sqrt{2}\vartheta^{2}\,A^{a}_{\mu}U^{a}_{\mu}
\end{equation}
does not renormalize, namely
\begin{equation}
(\vartheta^{2}A^{a}_{\mu}U^{a}_{\mu})_0=\vartheta^{2}A^{a}_{\mu}U^{a}_{\mu}\,,
\end{equation}
which implies that the mass parameter $\vartheta^2$ enjoys the following non-renormalization properties
\begin{equation}
Z_{\vartheta^{2}}=Z^{-1}_{A}\,, \label{nonren}
\end{equation}
and $a_2=-2a_1$ in eqs.\eqref{Zss}, meaning that its renormalization factor $Z_{\vartheta^{2}}$ is not an independent quantity.\\\\Let us first pay attention to the two-point 1PI Green's function of the gauge field $A^{a}_{\mu}$ at one-loop order\footnote{Due to the mirror symmetry, eq.\eqref{mirror}, equivalent results can be obtained for the replica field $U^{a}_{\mu}$.}. In this case, a counterterm like $\vartheta^{2}A^{a}_{\mu}A^{a}_{\mu}$ could arise, for example,  from a  tadpole diagram. This diagram is related to the following momentum integral
\begin{equation}
I=\int\frac{d^{d}q}{(2\pi)^{d}}\frac{q^{2}}{q^{4}+2\vartheta^{4}}\,,
\end{equation}
where $d=4-\varepsilon$ and where  use of the dimensional regularization is understood. Noticing that the Gribov propagator displays the following property
\begin{equation}
\frac{q^2}{q^4+2\vartheta^4} = \frac{1}{q^2}- \frac{2\vartheta^4}{q^2(q^4+2\vartheta^4)} \;, \label{gribdec}
\end{equation}
it follows that the integral $I$ can be rewritten as
\begin{equation}
I=\int\frac{d^{d}q}{(2\pi)^{d}}\frac{1}{q^{2}}-
2\vartheta^{4}\int\frac{d^{4}q}{(2\pi)^{4}}\frac{1}{q^{2}(q^{4}+2\vartheta^{4})}\,. \label{tadpole}
\end{equation}
The first term in expression \eqref{tadpole} vanishes  in dimensional regularization, while  the second integral is convergent in the ultraviolet region by power counting in $d=4$. The same argument is straightforwardly  applied to the other Feynman diagrams which contribute at one-loop order to the the two-point 1PI Green's function of the gauge field $A^{a}_{\mu}$.  Thus, one can state  that  no divergent counterterms of the form $\vartheta^{2}A^{a}_{\mu}A^{a}_{\mu}$ show up at one-loop order, in agreement with the general expression of the counterterm given in eq.\eqref{countert}. Moreover, a similar analysis can be performed for  the two-point mixed $A$-$U$ Green's function in order to prove that no divergent  counterterms of the type  $\vartheta^{2}A^{a}_{\mu}U^{a}_{\mu}$ are needed. In fact, the relevant one-loop Feynman integral in this case is of the type
\begin{equation}
I(k)=\int\frac{d^{d}q}{(2\pi)^{d}}\frac{-i\sqrt{2}\vartheta^{2}}{q^{4}+2\vartheta^{4}}\frac{-i\sqrt{2}\vartheta^{2}}{(q-k)^{4}+2\vartheta^{4}}\,,
\end{equation}
which is ultraviolet convergent by power counting. It is not difficult now, by making use of the decomposition \eqref{gribdec}, to extend these considerations to higher loop orders,  so as to conclude that no counterterms proportional to the mass parameter $\vartheta^2$ are in fact needed in order to renormalize the theory, as expressed by equation \eqref{nonren}.

%%%%%%%%%%%%%%%%%%%%%%%%%%%%%%%%%%%%
\subsection{Renormalizability of the refined version of the replica model}
%%%%%%%%%%%%%%%%%%%%%%%%%%%%%%%%%%%%
As already mentioned, the replica model, eq.\eqref{model}, admits a refined version \cite{Sorella:2010it} yielding a gluon propagator which does not vanish at the origin in momentum space, eq.\eqref{dec-prop}. The refined version of the replica model is obtained by adding to the action $S_{replica}$ of eq.\eqref{model} the following mass term
\begin{equation}
S_m=\int d^{4}x\,\frac{m^{2}}{2}(A^{a}_{\mu}A^{a}_{\mu}+U^{a}_{\mu}U^{a}_{\mu})\,, \label{masst}
\end{equation}
which turns out to be BRST invariant on-shell, {\it i.e.}
\begin{equation}
sS_m = m^2 \int d^4x \left( c^a \frac{\delta S_{ref}}{\delta b^a} + \omega^a \frac{\delta S_{ref}}{\delta {\bar b}^a} \right)  \;, \label{on-shell}
\end{equation}
where $S_{ref}$ stands for the refined action
\begin{equation}
S_{ref} = S_{replica} + S_m \;. \label{ref-act}
\end{equation}
The mass term, eq.\eqref{masst}, modifies the behavior of the propagators according to
\begin{eqnarray}
\langle A^{a}_{\mu}(k)A^{b}_{\nu}(-k) \rangle &=&\frac{k^{2}+m^{2}}{(k^{2}+m^{2})^{2}+2\vartheta^{4}}\delta^{ab}
\Bigl(\delta_{\mu\nu}-\frac{k_{\mu}k_{\nu}}{k^{2}}\Bigr)\,,\nonumber\\
\langle U^{a}_{\mu}(k)U^{b}_{\nu}(-k) \rangle &=&\frac{k^{2}+m^{2}}{(k^{2}+m^{2})^{2}+2\vartheta^{4}}\delta^{ab}
\Bigl(\delta_{\mu\nu}-\frac{k_{\mu}k_{\nu}}{k^{2}}\Bigr)\,,\nonumber\\
\langle A^{a}_{\mu}(k)U^{b}_{\nu}(-k) \rangle &=&\frac{-i\sqrt{2}\vartheta^{2}}{(k^{2}+m^{2})^{2}+2\vartheta^{4}}\delta^{ab}
\Bigl(\delta_{\mu\nu}-\frac{k_{\mu}k_{\nu}}{k^{2}}\Bigr)\,.
\end{eqnarray}
We notice that $S_m$ is left invariant by the discrete mirror symmetry \eqref{mirror} as well as by the rigid symmetries of eqs.\eqref{rig} and \eqref{rig2}.  As the replica model, also the refined action $S_{ref}$ turns out to be renormalizable. To that purpose, we follow the previous set up and embed the mass term $S_m$ into an extended action by means of a BRST doublet of external sources $(\lambda, \sigma)$
\begin{equation}
s\lambda=\sigma\,,\qquad s\sigma=0\,. \label{ls}
\end{equation}
It is thus easily verified that the expression $S_m$ is recovered from the BRST invariant term
\begin{eqnarray}
S_{\sigma}&=&s\int d^{4}x\,\Bigl(\lambda(A^{a}_{\mu}A^{a}_{\mu}+U^{a}_{\mu}U^{a}_{\mu})-\frac{\xi}{2}\,\lambda\sigma\Bigr)\nonumber\\
&=&\int d^{4}x\,\Bigl(\frac{1}{2}\sigma(A^{a}_{\mu}A^{a}_{\mu}+U^{a}_{\mu}U^{a}_{\mu})
+\lambda(A^{a}_{\mu}\,\partial_{\mu}c^{a}+U^{a}_{\mu}\,\partial_{\mu}\omega^{a})
+\frac{\xi}{2}\,\sigma^{2}\Bigr)\,, \label{actsigma}
\end{eqnarray}
when the sources $(\lambda, \sigma)$ attain the values
\begin{equation}
\sigma\bigl|_{\mathrm{phys}}=m^{2}\,,\qquad \lambda\bigl|_{\mathrm{phys}}=0\,.
\end{equation}
The last term of the r.h.s of expression \eqref{actsigma} is allowed by power counting, with  $\xi$ being a constant parameter. For the complete extended action accounting for the mass term $S_M$, we write
\begin{equation}
\Sigma'=\Sigma+S_{\sigma}\,,
\end{equation}
with $\Sigma$ given in eq.\eqref{Sigma}. Besides the Ward identities \eqref{gfix}--\eqref{sl2r}, the extended action
$\Sigma'$ enjoys an additional symmetry
\begin{equation}
\int d^{4}x\,\biggl(\frac{\delta\Sigma'}{\delta\lambda}
-ic^{a}\frac{\delta\Sigma'}{\delta{b}^{a}}
-i\omega^{a}\frac{\delta\Sigma'}{\delta{\bar{b}}^{a}}\biggr)=0\,.
\end{equation}
For the most general invariant counterterm we have now
\begin{equation}
\Sigma'_{c}=\Sigma_{c}+\int d^{4}x\,\Bigl(\frac{a_1}{2}\,\sigma(A^{a}_{\mu}A^{a}_{\mu}+U^{a}_{\mu}U^{a}_{\mu})
+\frac{a_{4}\xi}{2}\,\sigma^{2}\Bigr)\,,
\end{equation}
where $\Sigma_{c}$ is given in expression  \eqref{countert} and where $a_4$ stands for a free coefficient.
This counterterm can be reabsorbed in the action $\Sigma'$ by redefining the sources $(\sigma,\lambda)$ and the parameter $\xi$ as
\begin{equation}
\sigma_{0}=Z_gZ_A^{-1/2}\,\sigma\,,\qquad\lambda_{0}=Z_{g}^{1/2}Z_{A}^{-1/4}\,\lambda\,,\qquad
\xi_{0}=[1+\epsilon\,(2a_0+2a_1+a_4)]\,\xi\,,
\end{equation}
in addition to the previous redefinitions \eqref{Zs} and \eqref{Zss}.
 This completes the proof of the renormalizability of the refined version of the replica model.

\section{Construction of BRST invariant local composite operators with the quantum numbers $J^{PC} =0^{++}, 2^{++}, 0^{-+}$ and their $i$-particles content} \label{operators}
Glueballs are colorless composite gluon states  which are classified according to the values of the angular momentum $J$, parity $P$ and charge conjugation $C$. Within a quantum field theory framework, glueball states are constructed by means of suitable gauge invariant local composite operators carrying the quantum numbers $J^{PC}$ \cite{Mathieu:2008me,Jaffe:1985qp,DanMartin}. The physical properties of the glueballs, {\it i.e.} masses, decay properties, etc., are thus encoded in the analytic properties of the correlation functions of the corresponding composite operators. \\\\In what follows we shall look  at the construction of BRST local invariant composite operators describing the lightest glueball states, $J^{PC}=0^{++}, 2^{++}, 0^{-+}$, and whose correlation functions do have good analyticity properties. As already mentioned, see eqs.\eqref{iop} and \eqref{id}, this requirement can be achieved by looking at local operators which display an $i$-particles content. For the states with the quantum numbers  $J^{PC} =0^{++}, 0^{-+}$ these operators are easily identified \cite{Sorella:2010it} and given by
\begin{eqnarray}
{\cal O}_{0^{++}}(x) & =  &  \frac{1}{2}  \left(   F^a_{\mu\nu}(x)F^a_{\mu\nu}(x) - U^a_{\mu\nu}(x)U^a_{\mu\nu}(x)  \right)  \;, \label{o++} \\
{\cal O}_{0^{-+}}(x) & = & \frac{1}{2} \varepsilon_{\mu\nu\rho\sigma} \left( F^a_{\mu\nu}(x) F^a_{\rho\sigma}(x)- U^a_{\mu\nu}(x) U^a_{\rho\sigma}(x)\right)   \;.\label{o-+}
\end{eqnarray}
The case of the state $2^{++}$ requires more care. Usually, the $2^{++}$ state is associated with the energy-momentum tensor which, in the present context, does not display the required $i$-particles content.  Nevertheless, we can construct a local BRST invariant operator which creates a  pure $2^{++}$ state  by demanding that it is  symmetric, traceless and conserved. It has to be a symmetric tensor in order to have $J=2$.  It has to be traceless, otherwise its  trace would give rise to a state with the quantum numbers of the scalar glueball $0^{++}$. Finally, it has to be  conserved due to the fact that  its divergence would be associated with a vector glueball state. To construct such an operator we follow the procedure already employed in  \cite{DanMartin}, and we consider the local invariant operator
\begin{eqnarray}
\left[ {\cal O}_{2^{++}}(x) \right]_{\mu\nu}  =  \left(P_{\mu\alpha}P_{\nu\beta} - \frac 13 P_{\mu\nu}P_{\alpha\beta}\right) \left(   F^a_{\alpha\sigma}(x)F^a_{\beta\sigma}(x) - U^a_{\alpha\sigma}(x)U^a_{\beta\sigma}(x)\right), \label{2++}
\end{eqnarray}
where $P_{\mu\nu} \equiv \delta_{\mu\nu}\partial^2 - \partial_{\mu}\partial_{\nu}$ is the transverse projector.
Expression \eqref{2++} has the right properties for a pure $2^{++}$ state. It is in fact symmetric, traceless and conserved. Let us also show that, as required, all three operators, eqs.\eqref{o++}, \eqref{o-+}, \eqref{2++}, have an $i$-particles content.  \\\\Introducing the $i$-particles field variables, eqs.\eqref{ip},
\begin{eqnarray}
A^a_{\mu} & = \frac{1}{\sqrt{2}} \left( \lambda^a_{\mu} + \eta^a_{\mu}\right)\\
U^a_{\mu} & = \frac{1}{\sqrt{2}} \left( \lambda^a_{\mu} - \eta^a_{\mu}\right)\;,  \label{ifield}
\end{eqnarray}
it is easily verified that
\begin{eqnarray}
{\cal O}_{0^{++}}(x) & = &  \lambda^a_{\mu\nu}(x)\eta^a_{\mu\nu}(x) + {\ }{ higher \; order \; terms}, \label{io++} \\
\left[ {\cal O}_{2^{++}}(x) \right]_{\mu\nu} & = &  \left(P_{\mu\alpha}P_{\nu\beta} - \frac 13 P_{\mu\nu}P_{\alpha\beta}\right) \left(   \lambda^a_{\alpha\sigma}(x)\eta^a_{\beta\sigma}(x) + \eta^a_{\alpha\sigma}(x)\lambda^a_{\beta\sigma}(x)\right) + {\ }{ higher \; order \; terms},  {\ }{\ }{\ }{\ }{\ }{\ }\label{i2++}\\
{\cal O}_{0^{-+}}(x) & = & \frac{1}{2} \varepsilon_{\mu\nu\rho\sigma} \left( \lambda^a_{\mu\nu}(x) \eta^a_{\rho\sigma}(x)+ \eta^a_{\mu\nu}(x) \lambda^a_{\rho\sigma}(x)\right)  + {\ }{ higher \; order \; terms}, \label{io-+}
\end{eqnarray}
where $(\lambda^a_{\mu\nu}, \eta^a_{\mu\nu})$ stand for the quantities
\begin{eqnarray}
\lambda^a_{\mu\nu} & = \partial_{\mu}\lambda^a_{\nu} - \partial_{\nu}\lambda^a_{\mu}\\
\eta^a_{\mu\nu} & = \partial_{\mu}\eta^a_{\nu} - \partial_{\nu}\eta^a_{\mu}\;,  \label{ifieldstrength1}
\end{eqnarray}
and where higher order terms in the fields have been neglected as they will not enter the evaluation of the correlation functions at one-loop.

\subsection{Obtaining the K\"all\'{e}n-Lehmann spectral representation for the one-loop correlation functions}
At one-loop order, the two-point correlation functions of the glueball operators, eqs.\eqref{io++}, \eqref{i2++},\eqref{io-+},  take the general form
\begin{equation}
\langle {\cal O}_i(k) {\cal O}_i(-k) \rangle \Big|_{1-loop} = \int \frac{d^4p}{(2\pi)^4}\; \frac{1}{(k-p)^2+i\sqrt{2}\vartheta^2} \frac{1}{p^2-i\sqrt{2}\vartheta^2} f_i(p,k-p) \;. \label{1-loop}
\end{equation}
where $f_i(p,k-p)$, $i=0^{++}, 2^{++}, 0^{-+}$,  are polynomials  in the scalar products of the momenta $(k,p)$. There are several ways to show that  expressions of the form \eqref{1-loop}  do exhibit in fact a K\"all\'{e}n-Lehmann spectral representation, namely
\begin{equation}
\langle {\cal O}_i(k) {\cal O}_i(-k) \rangle \Big|_{1-loop}= \int^{\infty}_{0} d\tau \; \frac{\rho_i^{rep}(\tau)}{\tau + k^2}\;. \label{klrep}
\end{equation}
One possibility is to follow the procedure employed in \cite{Baulieu:2009ha}, relying on the use of Feynman parameters and of dimensional regularization. Here, we shall follow another route and establish the spectral representation by making use of the Cutkosky's rules in Minkowski space. At the end  we shall perform  an analytic continuation to complex masses in Euclidean  space. We shall explicitly see that the results obtained this way turn out to coincide in fact with those of \cite{Baulieu:2009ha}. \\\\Cutkosky's rules enable us to obtain the discontinuity of Feynman amplituted across the cuts. For that, each Feynman propagator corresponding to a cut internal line is put on-shell, according to
\begin{equation}
\label{cutk-cutrules}
 \frac{1}{p^2-m^2}\rightarrow 2\pi \theta(p^0)\delta(p^2-m^2) \;.
\end{equation}
Thus, for the discontinuity of the one-loop Feynman integral in Minkowski space
\begin{equation}
\int \frac{d^4p}{(2\pi)^4}\; \frac{1}{(k-p)^2-m^2_1} \frac{1}{p^2-m^2_2} f_i(p,k-p) \;, \label{1-loopm}
\end{equation}
we get
\begin{align}
Im {\cal F}_i = \int \frac{d^4p}{(2\pi)^4}\; (2\pi)^2 \theta((k-p)^0)\delta((k-p)^2-m_1^2)\theta(p^0)\delta(p^2-m_2^2)f_i(p, (k-p)) \;. \label{cutk}
\end{align}
We are ultimately interested in the case where the masses involved are complex, {\it i.e.} $m_1^2 \rightarrow m^2_{\lambda} = i\sqrt{2}\vartheta^2$ and $m_2^2 \rightarrow m^2_{\eta} = -i\sqrt{2}\vartheta^2$ with the momenta defined in Euclidean space. We then evaluate expression (\ref{cutk}) as a function of real masses and Minkowsky momenta and we analytically continue the resulting expression to  complex masses in Euclidean momentum space. This leads to the following prescription for the expression of the spectral functions entering expression \eqref{klrep}
\begin{align}
\rho_{i}^{rep}(\tau, m_{\lambda},m_{\eta}) = \rho^{Mink}_i(\tau,m_1,m_2)\Bigr\rvert_{m_1=m_{\lambda}; m_2=m_{\eta}}  \;, \label{spcfunc}
\end{align}
where $\rho^{Mink}_i(\tau,m_1,m_2)$ in the right hand side is obtained from expression \eqref{cutk}, namely
\begin{equation}
\rho^{Mink}_i = \frac{1}{\pi}  Im {\cal F}_i  \;. \label{rhoImF}
\end{equation}
\\\\The integral \eqref{cutk} is readily evaluated, and we find the spectral representation at one-loop order for each of the two-point correlation function of the glueball operators:
\begin{eqnarray}
\langle {\cal O}_{0^{++}}(k) {\cal O}_{0^{++}}(-k) \rangle \Big|_{1-loop}  & = & \frac{2(N^2-1)}{8\pi^2} \int_{2\sqrt{2}\vartheta^2}^{\infty} d\tau
\frac{ \sqrt{1-\frac{8\vartheta^4}{\tau^2}}}{\tau+k^2}\left(\frac{8\vartheta^4}{2}+\tau^2\right)
\;, \label{klrepo++} \\[5mm]
\langle \left[ {\cal O}_{2^{++}}(k) \right]_{\mu\nu} \left[ {\cal O}_{2^{++}}(-k) \right]_{\mu\nu} \rangle \Big|_{1-loop}  & = & \frac{\frac 43 (N^2-1)}{8\pi^2} \int_{2\sqrt{2}\vartheta^2}^{\infty} d\tau
\frac{ \sqrt{1-\frac{8\vartheta^4}{\tau^2}}}{\tau+k^2}\left(\frac 78 (8\vartheta^4)^2 \tau^2 + 2(8\vartheta^4)\tau^4 + \frac 32 \tau^6\right)  \nonumber \\
\label{klrep2++} \\[5mm]
\langle {\cal O}_{0^{-+}}(k) {\cal O}_{0^{-+}}(-k) \rangle \Big|_{1-loop} & = &
 \frac{8(N^2-1)}{8\pi^2} \int_{2\sqrt{2}\vartheta^2}^{\infty} d\tau
\frac{ \sqrt{1-\frac{8\vartheta^4}{\tau^2}}}{\tau+k^2}\left(\tau^2-8\vartheta^4\right)
\;. \label{klrepo-+}
\end{eqnarray}
where the threshold is found to be given by the expression $2\sqrt{2}\vartheta^2 = (m_{\lambda}+m_{\eta})^2=(2^{\frac 14}e^{i\frac{\pi}{4}}\vartheta + 2^{\frac 14}e^{i\frac{-\pi}{4}}\vartheta)^2$. Notice that all spectral densities $\rho_i^{rep}$, $i=0^{++}, 2^{++}, 0^{-+}$ are positive in the range of integration. It is also worth pointing out  that expressions \eqref{klrepo++} and \eqref{klrepo-+}  coincide with those already reported in \cite{Baulieu:2009ha}.

%%%%%%%%%%%%%%%%%%%%%%%%%%%%%%%
\section{Establishing a SVZ-type sum rules. An exercise on the spectrum of the glueballs }
%%%%%%%%%%%%%%%%%%%%%%%%%%%%%%%
Having identified a good set of BRST invariant local composite operators with quantum numbers $J^{PC} =0^{++}, 2^{++}, 0^{-+}$ and with positive spectral densities, we might attempt at  achieving a qualitative preliminary analysis of the ratios of the three masses $m^2_{0^{++}}, m^2_{2^{++}}, m^2_{0^{-+}}$, in order to obtain a first indication of their location. To that purpose, we shall work out a kind of phenomenological SVZ-type sum rules containing  a free coefficient $a$ which parametrizes our lack of knowledge of the true physical spectral functions.  Further, an expression for the masses of the three glueballs  $0^{++}, 2^{++}, 0^{-+}$ as a function of the coefficient $a$ is obtained by making use of the  Borel transformation. In the spirit of the SVZ sum rules, we shall  look thus at the existence of an interval for the parameter $a$, for which the location of the three masses  $m^2_{0^{++}}, m^2_{2^{++}}, m^2_{0^{-+}}$ is in agreement with the lattice data, see ref.\cite{Mathieu:2008me}, according to which the lightest state is the $0^{++}$, followed by the $2^{++}$, the $0^{-+}$ being  the heaviest one.  The existence of such an interval will be taken  as a first  encouraging  evidence for a more complete  quantitative analysis. Let us thus proceed by describing how our phenomenological SVZ-type sum rules are worked out. \\\\As it is customary in the SVZ approach to QCD \cite{Colangelo:2000dp}, we start by considering the two-point correlation functions
\begin{equation}
\Pi_i(q^2) = \int d^4x \, e^{iqx} \langle O_i(x) O_i(0) \rangle \;, \label{cr1}
\end{equation}
where $O_i$, $i=0^{++}, 2^{++}, 0^{-+}$, stand for the local composite gauge invariant operators which generate glueball states with quantum numbers $J^{PC} =0^{++}, 2^{++}, 0^{-+}$. \\\\On physical grounds, a truly nonperturbative evaluation of $\Pi_i(q^2)$ would enable us to write an exact K\"all\'{e}n-Lehmann spectral representation
\begin{equation}
\Pi_i(q^2) = \frac{1}{\pi} \int_0^{\infty} d\tau \,\frac{Im \Pi_i(\tau)}{\tau+q^2}  \;, \label{kl}
\end{equation}
which is expected to follow from the unitarity and analyticity properties of the underlying nonperturbative theory\footnote{We remind here that, in some cases,  the spectral representation, eq.\eqref{cr1}, might require appropriate subtraction terms in order to ensure convergence. These terms are not written down, as they will be removed once the Borel transformation will be taken.}. We thus proceed by employing a one-resonance parametrization for $Im \Pi_i(\tau)$ \cite{Narison:1996fm}, namely
\begin{equation}
\frac{Im \Pi_i(\tau)}{\pi} = {\cal R}_i\, \delta(\tau-m^2_i) + \theta(\tau-\tau_0^i) \rho_i^{phys}(\tau) \;, \label{spdec}
\end{equation}
where $m^2_i$ denotes the glueball mass in the $i$-{\it th} channel, $\tau_0^i$ the  threshold for the continuum part of the spectrum and $\rho_i^{phys}(\tau)$ the corresponding positive spectral density. Of course, the real values of ${\cal R}_i$, $m^2_i$, $\tau_0^i$ and of the spectral density $\rho_i^{phys}(\tau)$ are unknown. So far, the best which can be done is computing the correlation functions \eqref{cr1} by trying to encode as much nonperturbative effects as possible, as summarized by the following equation
\begin{equation}
\frac{ {\cal R}_i}{q^2+m^2_i}  + \int_{\tau_0^i}^{\infty} d\tau\, \frac{\rho_i^{phys}(\tau)}{\tau+q^2} =  \Pi_i^{np} \;, \label{sr}
\end{equation}
where $\Pi_i^{np}$ stands for the expression of the correlation function \eqref{cr1} which one has been able to evaluate in practice. Expression \eqref{sr} establishes the so-called sum rules, enabling us to give estimates of the glueball masses in terms of the nonperturbative parameters present in $\Pi_i^{np}$.  \\\\As shown in their seminal work \cite{Shifman:1978bx,Shifman:1978by,Shifman:1979if,Novikov:1981xi}, this approach has been proven to be rather successful in order to obtain estimates of  hadronic masses in QCD.\\\\Here, we shall not follow the standard path of the SVZ sum rules. We remind here that in the SVZ sum rules approach the quantity $\Pi_i^{np}$ in the right hand side of eq.\eqref{sr} is obtained through several steps. First, one starts  from the  usual QCD Lagrangian\footnote{{\it i.e.} $\int d^4x \left( \frac{1}{4}F^2 +{\bar \psi} \gamma_\mu D_\mu \psi  +m {\bar \psi}\psi \right)$.} which is employed to evaluate the perturbative contributions till a certain order. Further, non-perturbative contributions are included by the introduction of a suitable set of condensates obtained via operator product expansion (OPE). Finally, additional non-perturbative effects are obtained from instantons, see, for example,  \cite{Narison:1996fm,Forkel:2003mk}  for a detailed account. In the present work, we follow a slightly different route. The non-perturbative effects are already encoded in the starting action by means of a non-perturbative mass parameter $\vartheta^2$ which accounts for the presence of the Gribov horizon. More precisely, we shall attempt  attempt at evaluating the right hand side of eq.\eqref{sr} by employing a trial nonperturbative action $S^{np}$ fulfilling the following requirements:

\begin{itemize}
\item $i)$ the action  $S^{np}$ exhibits the nonperturbative effects of the Gribov horizon, whose presence is parametrized by the Gribov mass parameter, here denoted by  $\vartheta^2$. Moreover, in the limit $\vartheta^2=0$, in which the Gribov horizon is removed, the action  $S^{np}(\vartheta^2)$ reduces to the usual perturbative Faddeev-Popov action $S_{FP}$, {\it i.e.}
\begin{equation}
S^{np}(\vartheta^2) \Bigl|_{\vartheta^2=0}= S_{FP}  \;, \label{fp}
\end{equation}
\item $ii)$ it accounts for gluon confinement. This means that the two-point correlation function of the elementary gluon field appearing in  $S^{np}(\vartheta^2)$ cannot be cast in the form of a spectral representation with positive spectral density, so that it cannot be interpreted as the propagator of a physical particle.
\item $iii)$ it is renormalizable, meaning that consistent  calculations can be worked out.
\item $iv)$ it enables us to introduce gauge invariant or, equivalently, BRST invariant local composite operators $O_i^{np}$ with the quantum numbers $J^{PC} =0^{++}, 2^{++}, 0^{-+}$, whose two-point correlation functions
\begin{equation}
\Pi_i^{np}(q^2,\vartheta^2) = \int d^4x \, e^{iqx} \langle O_i^{np}(x) O_i^{np}(0) \rangle \;, \label{cr2}
\end{equation}
exhibit the K\"all\'{e}n-Lehmann spectral representation, a feature which we shall consider here at one-loop order only, {\it i.e.}
\begin{equation}
\Pi_i^{np}(q^2,\vartheta^2) = \int_{\tau_0^{(i)np}(\vartheta^2)}^{\infty} d\tau \,\frac{\rho_i^{np}(\tau)}{\tau+q^2} + {\cal O}(\hbar^2) \;. \label{kl1}
\end{equation}
\end{itemize}
Requirements $i)$--$iv)$ look quite stringent. As such, they might provide a satisfactory set up in order to achieve a useful predictive expression for the correlation functions  $\Pi_i^{np}(q^2,\vartheta^2)$, eq.\eqref{cr2}. \\\\Two examples of field theories compatible with the requirements $i)$--$iv)$ are known so far. The first example is provided by the Gribov-Zwanziger action \cite{Gribov:1977wm,Zwanziger:1988jt,Zwanziger:1989mf} and its refined version  \cite{Dudal:2007cw,Dudal:2008sp}.  Though, till now, the item $iv)$ has not yet been completely settled \cite{Baulieu:2009ha}. The second model available is the  replica model discussed here. As already mentioned, this model can be considered as a useful renormalizable laboratory reproducing most of the features of the Gribov-Zwanziger theory. It contains a soft mass parameter $\vartheta^2$ which plays a role akin to that of the Gribov mass parameter, while allowing for the introduction of BRST invariant local composite operators with quantum numbers $J^{PC} =0^{++}, 2^{++}, 0^{-+}$, whose two point correlation functions exhibit the K\"all\'{e}n-Lehmann spectral representation at one-loop order, thus fulfilling item $iv)$.  It is worth mentioning that both models rely on the mechanism of the soft breaking of the BRST symmetry. In the following, the action of the replica model will be taken as our trial nonperturbative action, {\it i.e.}
\begin{equation}
S^{np}(\vartheta^2) = S_{replica} \;. \label{npaction}
\end{equation}
According to expression \eqref{sr}, we shall thus elaborate on the equation
\begin{equation}
\frac{ {\cal R}_i}{q^2+m^2_i}  + \int_{\tau_0^i}^{\infty} d\tau\, \frac{\rho_i^{phys}(\tau)}{\tau+q^2} = \int_{2\sqrt{2}  \vartheta^2}^{\infty} d\tau \,\frac{\rho_i^{rep}(\tau)}{\tau+q^2} + {\cal O}(\hbar^2)   \;, \label{fpc1}
\end{equation}
where $\rho_i^{rep}(\tau)$ are the one-loop spectral densities given in eqs.\eqref{klrepo++},\eqref{klrep2++},\eqref{klrepo-+}. \\\\As in the SVZ sum rules approach \cite{Colangelo:2000dp}, before going any further we need to  provide an estimate for the quantity
\begin{equation}
\int_{\tau_0^i}^{\infty} d\tau\, \frac{\rho_i^{phys}(\tau)}{\tau+q^2} \;, \label{est}
\end{equation}
appearing in eq.\eqref{fpc1}. Concerning the threshold ${\tau_0^i}$, we shall write
\begin{equation}
\tau_0^i = a 2 \sqrt{2}  \vartheta^2   \;, \qquad i=0^{++}, 2^{++}, 0^{-+} \;, \label{trsa}
\end{equation}
where $a >1$ stands for a free parameter accounting for the difference between the value of the physical threshold, $\tau_0^i$, and that which we have been able to evaluate in practice, {\it i.e.}: $2\sqrt{2}\vartheta^2$. One should notice that, in principle, the physical threshold $\tau_0^i$ could be different for each glueball correlation function. Instead, we are employing a unique parameter $a$ for all states $i=0^{++}, 2^{++}, 0^{-+}$, a choice which is motivated by the fact that three correlation functions \eqref{klrepo++},\eqref{klrep2++},\eqref{klrepo-+}  exhibit the same threshold.
Further, for what concerns the spectral density ${\rho_i^{phys}(\tau)}$, we shall proceed by assuming that, at the one-loop order approximation considered here, it can be well approximated by ${\rho_i^{rep}(\tau)}$, and that possible deviations start to show up at higher orders, {\it i.e.} we shall set
\begin{equation}
\rho_i^{phys}(\tau) = \rho_i^{rep}(\tau) + {\cal O}(\hbar^2) \;.  \label{sf1}
\end{equation}
Therefore, for the final form of the sum rules, we write
\begin{equation}
\frac{ {\cal R}_i}{q^2+m^2_i} = \int_{2\sqrt{2}\vartheta^2}^{2\sqrt{2}\,a \vartheta^2} d\tau \,\frac{\rho_i^{rep}(\tau,\vartheta^2)}{\tau+q^2}  + {\cal O}(\hbar^2)\;. \label{fsr1}
\end{equation}
Equation \eqref{fsr1}  will be employed for a qualitative discussion of the ratios of the masses of the lightest glueball. This will be done by looking at the existence of an interval for the phenomenological parameter $a$ for which the ratios of the glueball masses obtained through eq.\eqref{fsr1}  are in qualitative agreement with the available lattice data. Let us also underline that the necessary non-perturbative ingredient for a study of the spectrum  of the glueballs, even at the qualitative level, is encoded here in the presence of the parameter $\vartheta^2$ which, to some extent, accounts for the effects of the Gribov horizon.

\subsection{A formula for the glueball masses}
Expression \eqref{fsr1} enables us to extract a formula for the glueball masses. To that purpose, we shall first compute the Borel transformation of the first moment of the sum rules \eqref{fsr1}, namely
\begin{equation}
{\cal B}_M\left(  \frac{q^2 {\cal R}_i}{q^2+m^2_i} \right) = {\cal B}_M \left( q^2 \int_{2\sqrt{2}\vartheta^2 }^{2\sqrt{2}\,a \vartheta^2} d\tau \,\frac{\rho_i^{rep}(\tau)}{\tau+q^2}\right) \;, \label{borel1}
\end{equation}
where ${\cal B}_M$ stands for the Borel operator \cite{Mathieu:2008me,Colangelo:2000dp}
\begin{equation}
{\cal B}_M(f(q^2)) = {\rm lim}_{n,q^2 \rightarrow \infty} \frac{(-1)^n}{n!} (q^2)^{n+1} \left(\frac{d^nf(q^2)}{d^nq^2}\right) \Biggl|_{\frac{q^2}{n}=M^2= {\rm fixed}} \;, \label{borelop}
\end{equation}
and $M^2$ is the so called Borel mass \cite{Mathieu:2008me,Colangelo:2000dp}.\\\\From
\begin{equation}
{\cal B}_M \left(  \frac{1} {q^2+m^2_i} \right) = e^{-\frac{m^2_i}{M^2}} \;, \label{br1}
\end{equation}
one easily gets
\begin{equation}
{\cal R}_i m^2_i e^{-\frac{m^2_i}{M^2}} = \int_{2\sqrt{2} \vartheta^2}^{2\sqrt{2}\;a\vartheta^2} d\tau \,\tau \, \rho_i^{rep}(\tau) \,  e^{-\frac{\tau}{M^2}} \;. \label{borel2}
\end{equation}
Taking now the Borel transformation of eq.\eqref{fsr1}
\begin{equation}
{\cal R}_i  e^{-\frac{m^2_i}{M^2}} =  \int_{2\sqrt{2} \vartheta^2}^{2\sqrt{2}\; a \vartheta^2} d\tau  \, \rho_i^{rep}(\tau) \,  e^{-\frac{\tau}{M^2}}  \;, \label{borel3}
\end{equation}
and evaluating the ratio between eq.\eqref{borel2} and eq.\eqref{borel3}, we get the mass formula  we are looking for
\begin{equation}
m^2_i = \frac{  \int_{2\sqrt{2} \vartheta^2}^{2\sqrt{2} \, a \vartheta^2} d\tau \,\tau \, \rho_i^{rep}(\tau) \,  e^{-\frac{\tau}{M^2}}    }{  \int_{2\sqrt{2} \vartheta^2}^{2\sqrt{2} \, a \vartheta^2} d\tau  \, \rho_i^{rep}(\tau) \,  e^{-\frac{\tau}{M^2}} }  \;. \label{massg}
\end{equation}
An interesting aspect of expression \eqref{massg} is that the  mass $m^2_i$ exhibits an explicit dependence from both threshold $2\sqrt{2}\vartheta^2$ and spectral density  ${\rho_i^{rep}(\tau)}$.

\subsection{A qualitative analysis of the ratios of the glueball masses $m^2_{0^{++}}, m^2_{2^{++}}, m^2_{0^{-+}}$}

We have now all ingredients for a qualitative analysis of the glueball masses $m^2_{0^{++}}, m^2_{2^{++}}, m^2_{0^{-+}}$. Making use of expressions  \eqref{klrepo++},\eqref{klrep2++},\eqref{klrepo-+}, equation \eqref{massg} can be written in explicit form giving
\begin{eqnarray}
m^2_{0^{++}} & =&  \frac{ \int_{2\sqrt{2}\vartheta^2}^{2\sqrt{2}\, a\vartheta^2} d\tau \sqrt{\tau^2-8\vartheta^4}\, \left(\frac{8\vartheta^4}{2}+\tau^2\right)\, e^{-\frac{\tau}{M^2}}      } { \int_{2\sqrt{2}\vartheta^2}^{2\sqrt{2}\, a\vartheta^2} d\tau \frac{\sqrt{\tau^2-8\vartheta^4}\, \left(\frac{8\vartheta^4}{2}+\tau^2\right)} {\tau}\, e^{-\frac{\tau}{M^2}}  }\;, \label{m0++} \\[5mm]
m^2_{2^{++}} & = &  \frac{ \int_{2\sqrt{2}\vartheta^2}^{2\sqrt{2}\, a\vartheta^2} d\tau \sqrt{\tau^2-8\vartheta^4}\, \left(\frac 78 (8\vartheta^4)^2 \tau^2 + 2(8\vartheta^4)\tau^4 + \frac 32 \tau^6\right)\, e^{-\frac{\tau}{M^2}}      } { \int_{2\sqrt{2}\vartheta^2}^{2\sqrt{2}\, a\vartheta^2} d\tau \frac{\sqrt{\tau^2-8\vartheta^4}\, \left(\frac 78 (8\vartheta^4)^2 \tau^2 + 2(8\vartheta^4)\tau^4 + \frac 32 \tau^6\right)} {\tau}\, e^{-\frac{\tau}{M^2}}  } \;, \label{m2++} \\[5mm]
m^2_{0^{-+}} & = &  \frac{ \int_{2\sqrt{2}\vartheta^2}^{2\sqrt{2}\, a\vartheta^2} d\tau \left(\tau^2-8\vartheta^4\right)^{3/2} \, e^{-\frac{\tau}{M^2}}      } { \int_{2\sqrt{2}\vartheta^2}^{2\sqrt{2}\, a\vartheta^2} d\tau \frac{\left(\tau^2-8\vartheta^4\right)^{3/2}} {\tau}\, e^{-\frac{\tau}{M^2}}   } \;. \label{m0-+}
\end{eqnarray}
It is possible to rewrite these expressions in such a way to make all the masses proportional to the mass parameter $\vartheta^2$. We introduce the variable $t = \frac{\tau}{2\sqrt{2}\vartheta^2}$ and the parameter $p = \frac{2\sqrt{2}\vartheta^2}{M^2}$. The resulting expressions read
\begin{eqnarray}
m^2_{0^{++}}(a,p) & =& 2\sqrt{2}\vartheta^2 \frac{ \int_{1}^{a} dt \sqrt{t^2-1}\, (\frac 12 + t^2)\, e^{-pt}    } { \int_{1}^{a} dt \frac{\sqrt{t^2-1}\, (\frac 12 + t^2)} {t}\, e^{-pt}  } \;, \label{m0++a} \\[5mm]
m^2_{2^{++}}(a,p) & =& 2\sqrt{2}\vartheta^2 \frac{ \int_{1}^{a} dt \sqrt{t^2-1}\, (\frac 78 t^2 +2t^4 + \frac 32 t^6)\, e^{-pt}       } { \int_{1}^{a} dt \frac{\sqrt{t^2-1}\, (\frac 78 t^2 +2t^4 + \frac 32 t^6)} {t}\, e^{-pt}  } \;, \label{m2++a} \\[5mm]
m^2_{0^{-+}}(a,p) & = &  2\sqrt{2}\vartheta^2\frac{ \int_{1}^{a} dt \left(t^2-1\right)^{3/2} \, e^{-pt}        } { \int_{1}^{a} dt \frac{\left(t^2-1\right)^{3/2}} {t}\, e^{-pt}   } \;. \label{m0-+a}
\end{eqnarray}
The masses become now functions of the parameters $a$ and $p$. The output of our results are shown in Fig.1, Fig.2 and Fig.3.  From Fig.1 one can see that, when the parameter $a$ belongs to the interval  $1<a < 1.8$, the masses of the three lightest states are in qualitative agreement with the available data, {\it i.e.} $m^2_{0^{++}} < m^2_{2^{++}} < m^2_{0^{-+}}$, a feature which holds for all values of $p$, as shown in Fig.2. and in the three dimensional plots of Fig.3. The existence of the interval $1<a < 1.8$ is regarded as a preliminary test towards a more quantitative analysis of the glueballs spectrum.
\begin{figure}[H]
\label{fig:0pp2pp0mp}
  \centering
  {\includegraphics[scale=1]{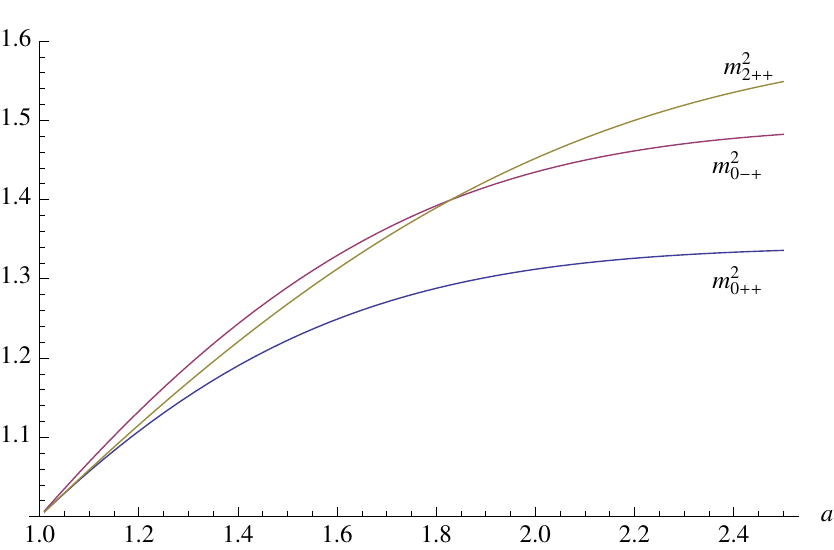}}
  \caption{Glueball masses as functions of the threshold parameter $a$ of eq.\eqref{trsa} for $p=5$, with $p=\frac{2\sqrt{2}\vartheta^2}{M^2}$, where $M$ is the Borel mass and $\vartheta^2$ the mass parameter of the starting action in eq.\eqref{model}}
\end{figure}

\begin{figure}[H]
\label{fig:0pp2pp0mp-p}
 \centering
 {\includegraphics[scale=1]{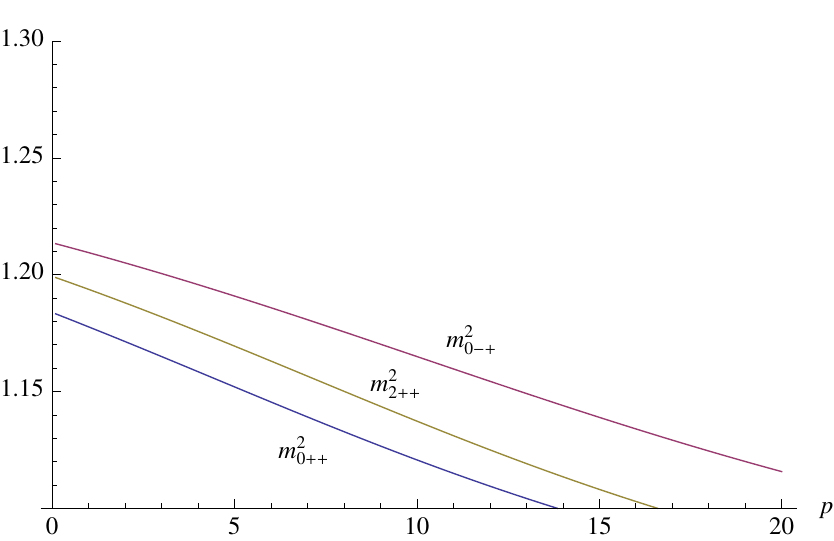}}
 \caption{Glueball masses as functions of $p$ for $a=1.3$}
\end{figure}

\begin{figure}[H]
  \centering
  \subfloat[$\frac{m^2_{0^{++}}}{m^2_{0^{-+}}}(a,p)$]{\label{fig:0pp-0mp}\includegraphics[scale=0.5]{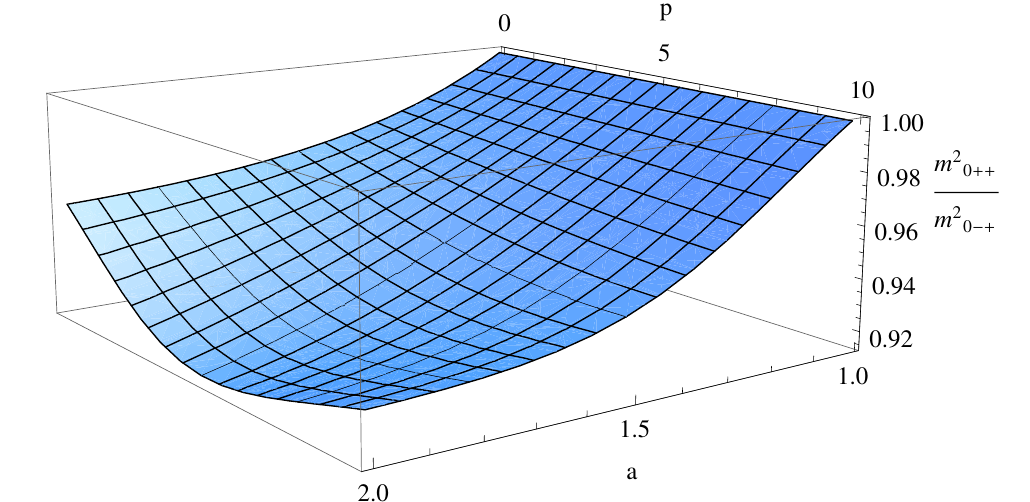}}\hspace{100pt}
  \subfloat[$\frac{m^2_{0^{++}}}{m^2_{2^{++}}}(a,p)$]{\label{fig:0pp-2pp}\includegraphics[scale=0.5]{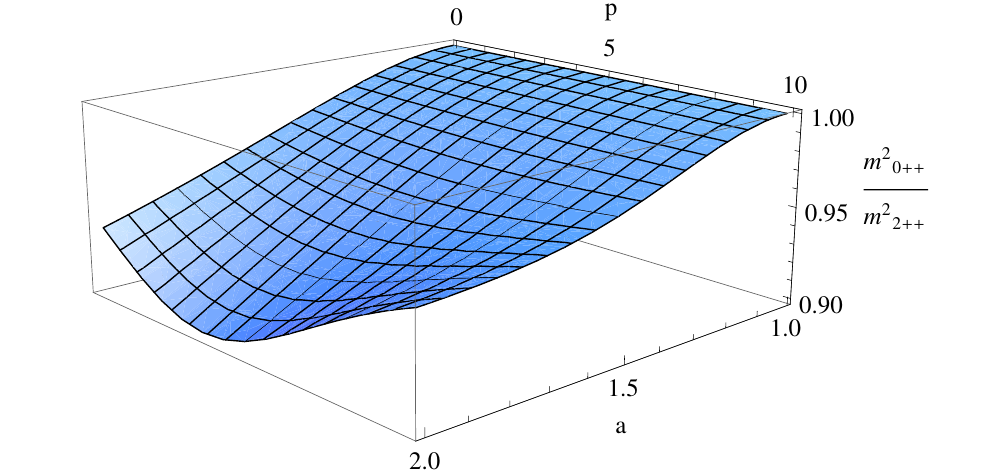}}\hspace{100pt}
  \subfloat[$\frac{m^2_{2^{++}}}{m^2_{0^{-+}}}(a,p)$]{\label{fig:2pp-0mp}\includegraphics[scale=0.5]{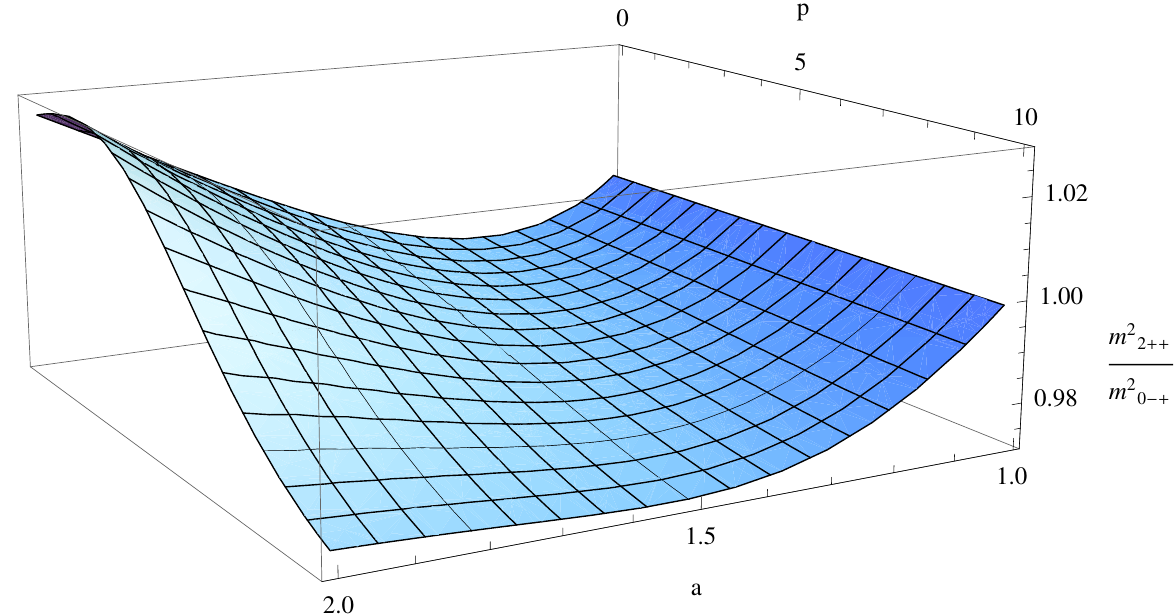}}
  \caption{Glueball mass ratios}
\end{figure}

\section{Conclusion}

In this work several aspects of the replica model introduced in \cite{Sorella:2010it} have been addressed. The relevance of this model relies on the fact that it reproduces several features of the Gribov-Zwanziger theory \cite{Gribov:1977wm,Zwanziger:1988jt,Zwanziger:1989mf}. As such, it can be employed as a useful laboratory to investigate, within the framework of the Euclidean quantum field theory, the consequences of having a confining gluon propagator of the Gribov type, eq.\eqref{paai}. \\\\We have shown that the replica model is a renormalizable theory, while enabling us to introduce suitable BRST invariant local composite operators with the quantum numbers of the lightest glueball states, $J^{PC} =0^{++}, 2^{++}, 0^{-+}$. The corresponding correlation functions have been evaluated at one-loop order and shown to display a spectral representation with positive spectral densities. \\\\A first  check of the fact that the nonperturbative effects related to the Gribov horizon might be useful for the investigation of the properties of the glueball spectrum has been achieved through a  phenomenological SVZ-type sum rules, eq.\eqref{fsr1}.  Although the analysis presented here is only at  the  qualitative level, it can open the road for a  more quantitative study in which the most recent available lattice data on the gluon two-point correlation function can be employed to get numerical estimates of the Gribov mass parameter and of the related dimension two condensates \cite{Dudal:2010tf}. This could enable us to obtain quantitative estimates for  the glueball masses $m^2_{0^{++}}$, $m^2_{2^{++}}$, $m^2_{0^{-+}}$, to be compared with the available data.\\\\Finally, although not reported, we mention that plots similar to those of Figures 1,2,3 can be obtained in the case of the decoupling propagator, eq.\eqref{dec-prop}, which also displays the $i$-particles decomposition
\begin{equation}
 \frac{k^{2}+m^{2}}{(k^{2}+m^{2})^{2}+2\vartheta^{4}} = \frac{1}{2} \left( \frac{1}{k^2+m^2+i\sqrt{2}\vartheta^2} +
 \frac{1}{k^2+m^2-i\sqrt{2}\vartheta^2} \right) \;. \label{i-dec}
\end{equation}

%%%%%%%%%%%%%%%%%%%%%%%%%%
\section*{Acknowledgments}
%%%%%%%%%%%%%%%%%%%%%%%%%%%
We thank David Dudal, Nele Vandersickel and Daniel Zwanziger for useful discussions and comments. \\\\The Conselho Nacional de Desenvolvimento Cient\'{\i}fico e
Tecnol\'{o}gico (CNPq-Brazil), the Faperj, Funda{\c{c}}{\~{a}}o de
Amparo {\`{a}} Pesquisa do Estado do Rio de Janeiro, the Latin
American Center for Physics (CLAF), the SR2-UERJ,  the
Coordena{\c{c}}{\~{a}}o de Aperfei{\c{c}}oamento de Pessoal de
N{\'{\i}}vel Superior (CAPES)  are gratefully acknowledged.

\end{document}